\documentclass[12pt]{article}
\usepackage{epsfig, amsmath, amssymb}
\usepackage{graphicx,psfrag} 
\newcommand{\be}{\begin{equation}}
\newcommand{\ee}{\end{equation}}
\newcommand{\bea}{\begin{eqnarray}}
\newcommand{\eea}{\end{eqnarray}}
\newcommand{\bA}{\begin{array}}
\newcommand{\eA}{\end{array}}
\newcommand{\bc}{\begin{center}}
\newcommand{\ec}{\end{center}}
\newcommand{\al}{\alpha}

\newcommand{\ra}{\rightarrow}
\newcommand{\del}{\partial}

\newcommand{\ie}{{\it i.e.}}
\newcommand{\eg}{{\it e.g.}}

\newcommand{\Nf}{${\cal N}{=}4$}

\newcommand{\No}{${\cal N}{=}1$}

\begin{document}


\begin{titlepage}
\vspace{30mm}

\bc

\hfill 
\\         [22mm]

{\Huge $AdS$ null deformations 
with inhomogeneities }
\vspace{16mm}

{\large K.~Narayan} \\
\vspace{3mm}
{\small \it Chennai Mathematical Institute, \\}
{\small \it SIPCOT IT Park, Siruseri 603103, India.\\}

\ec
\medskip
\vspace{40mm}

\begin{abstract}
We study $AdS\times X$ null deformations arising as near horizon
limits of D3-brane analogs of inhomogenous plane waves. Restricting to
normalizable deformations for the $AdS_5$ case, these generically
correspond in the dual field theory to SYM states with lightcone
momentum density $T_{++}$ varying spatially, the homogenous case
studied in arXiv:1202.5935 [hep-th] corresponding to uniform
$T_{++}$. All of these preserve some supersymmetry. Generically these
inhomogenous solutions exhibit analogs of horizons in the interior
where a timelike Killing vector becomes null. From the point of view
of $x^+$-dimensional reduction, the circle pinches off on these
horizon loci in the interior. We discuss similar inhomogenous
solutions with asymptotically Lifshitz boundary conditions, as well as
aspects of Lifshitz singularities in string constructions involving
$AdS$ null deformations. We also briefly discuss holographic
entanglement entropy for some of these.
\end{abstract}

\end{titlepage}

\newpage 
{\tiny 
\begin{tableofcontents}
\end{tableofcontents}
}


\section{Introduction}

Various fascinating explorations of strongly coupled quantum field
theories have been carried out using gauge/gravity duality
\cite{AdSCFT}, including, more recently, non-relativistic and
condensed matter systems \cite{AdSCondmatRev} with symmetries
typically smaller than anti de Sitter space. Several interesting
features of finite density systems can in fact be simulated in fairly
simple effective gravity models with cosmological constant and
vector/scalar matter sources. It is important to understand such
models in string theory: for one thing, it is expected that the
parameter space of string constructions is more constrained, and it
might be possible to track the stringy origins of the effective
parameters. Furthermore, a string/brane construction might 
suggest natural field theory duals to the gravity descriptions.

A remarkably simple family of such string realizations of some 
nonrelativistic systems involves null deformations of $AdS\times X$ 
spacetimes that arise in familiar brane constructions, of the form
\be\label{AdSnullLif0}
ds^2 = {1\over r^2} [-2dx^+dx^- + dx_i^2 + dr^2] + g_{++} (dx^+)^2 
+ d\Omega_S^2\ , 
\ee
where the metric component $g_{++}$ might be sourced by various fields.
For instance, spacetimes with $z=2$ Lifshitz scaling symmetry 
\cite{Kachru:2008yh,Taylor:2008tg} can be realized in string 
constructions via dimensional reduction along the $x^+$-direction of 
metrics (\ref{AdSnullLif0}) with non-normalizable deformations 
$g_{++}\sim {1\over r^2} r^2$ \cite{Balasubramanian:2010uk,Donos:2010tu} 
(see also \eg\ \cite{Gregory:2010gx}). Likewise, metrics conformal 
to Lifshitz spacetimes arise in holographic systems with nontrivial 
hyperscaling violation exponents encoded in the conformal factor: 
these have been discussed in \eg\ \cite{Ogawa:2011bz,Huijse:2011ef} 
motivating and clarifying connections with condensed matter systems 
(in particular pertaining to holographic entanglement entropy 
\cite{Ryu:2006bv} and Fermi surfaces): see also \cite{Dong:2012se} 
for various aspects of holography in this context. Some of these 
hyperscaling violating metrics can be realized by $x^+$-dimensional 
reduction of metrics (\ref{AdSnullLif0}) using normalizable 
deformations $g_{++}\sim {1\over r^2} r^4$ \cite{Narayan:2012hk}\ \
(see also \eg\ \cite{Dong:2012se,Kim:2012nb,Singh:2012un,Dey:2012tg,
Perlmutter:2012he,Ammon:2012je,Kulaxizi:2012gy} for other string 
realizations of such models, and \eg\ \cite{Goldstein:2009cv,Cadoni:2009xm,
Charmousis:2010zz,Perlmutter:2010qu,Gouteraux:2011ce,Bertoldi:2010ca,
Iizuka:2011hg,Iizuka:2012iv,Alishahiha:2012cm} for effective gravity 
models with vectors and scalars where these and related metrics 
arise, with associated physics). In particular, the $AdS_5$
normalizable null deformation after dimensional reduction
interestingly gives a spacetime with $\theta=1, d=2$, which exhibits
logarithmic violation of the area law of entanglement entropy in the
holographic context.

In this paper, we will explore the space of such $AdS\times X$ null
deformations in greater generality, allowing for possible
inhomogeneities, \ie\ with $g_{++}$ having spatial ($x_i$) dependence.
These are near horizon limits of D3-brane analogs of plane-waves with
possible inhomogeneities (see \eg\ \cite{Craps:2011sp} for a recent
review discussing plane-wave backgrounds \cite{Horowitz:1989bv} in the
context of cosmological singularities). Since $AdS_5\times S^5$ is
$\al'$-exact \cite{Kallosh:1998qs} as are plane wave spacetimes, these
$AdS$ null deformations are also likely $\al'$-exact string
backgrounds.  We will restrict in particular to static $AdS$
deformations that are normalizable near the boundary, \ie\
normalizable backgrounds, so they can be interpreted as states in the
dual super Yang-Mills theory with nontrivial lightcone momentum
density $T_{++}$ that might vary spatially (regarding $x^+$ as a
noncompact direction, sec. 2). In general, the structure of these
normalizable background solutions involves modes that grow in the
interior: this implies that for large families of solutions with
inhomogeneities, $g_{++}$ vanishes somewhere in the interior, even if
it is positive definite near the boundary. These $g_{++}=0$ loci are
akin to horizons, in the sense that a timelike Killing vector $\del_-$
becomes null.

Part of the motivation here is to understand the ``vicinity'' of 
the homogenous $AdS$ plane wave studied in \cite{Narayan:2012hk}.
In other words, we would like to explore ``nearby'' 
solutions, at least within the class describable as $AdS$ null
deformations. From the dual field theory point of view, the homogenous 
case has uniform lightcone momentum density $T_{++}\sim Q$, so that 
$T_{++}\sim Q+\epsilon f(x,y)$ with small $\epsilon$ (and localized 
near some $x_{i0}$) would constitute 
a ``small'' inhomogenous perturbation. Then we find that starting 
with a homogenous background $g_{++}$, turning on a ``small'' 
inhomogenous perturbation near the boundary (\ie\ $T_{++}$ as above) 
corresponds to a bulk spacetime which departs substantially
from the homogenous $AdS$ plane wave, due to the emergence of horizon
loci.

From the perspective of $x^+$-dimensional reduction (sec. 3) of such
backgrounds, it appears that the circle pinches off at the loci where
$g_{++}=0$, and new states emerge corresponding to string winding
modes that become light in the vicinity of these loci.  We also
explore asymptotically Lifshitz backgrounds with inhomogeneities, and
discuss Lifshitz singularities (arising from diverging tidal forces in
the interior) from the point of view of string constructions involving
$AdS$ null deformations.

Finally, we discuss holographic entanglement entropy briefly for 
these null deformed backgrounds (sec. 4) from the point of view 
of the higher dimensional description (\ie\ with $x^+$-noncompact).

Our discussion is primarily for the $AdS_5$ case arising from 
D3-branes: we also briefly analyse other $AdS_D$ null deformations 
of this sort (sec. 2), and expect similar features.

\section{$AdS_5$ null deformations with inhomogeneities}

We are considering spacetimes of the form (\ref{AdSnullLif0}), \ie\
\be\label{AdSnullLif}
ds^2 = {1\over r^2} [-2dx^+dx^- + dx_i^2 + dr^2] + g_{++} (dx^+)^2 
+ d\Omega_S^2\ ,
\ee
as solutions to IIB string theory (or supergravity), obtained by 
null deformations of the familiar near horizon geometry of a 
D3-brane stack. Our discussion will be mostly for $AdS_5\times S^5$ 
dual to 4-dim \Nf\ SYM theory, but the arguments can also be 
generalized to other super Yang-Mills theories dual to 
$AdS_5\times X^5$ spacetimes, with the 5-space $X^5$ being a 
Sasaki-Einstein base or equivalently the 6-dim space transverse to 
the D3-branes being Ricci-flat.
More generally, the 10-dim spacetime
\be\label{AdSplanewave}
ds^2 = Z^{-1/2} [-2dx^+dx^-+dx_i^2+N(x_i,x^m) (dx^+)^2] + Z^{1/2} dx^mdx^m\ ,
\ee
$Z(x^m)$ being harmonic in the transverse space, with the 
corresponding 5-form flux describing the stack of D3-branes with 
null deformation gives in the near horizon limit the metrics 
(\ref{AdSnullLif}) above for a single D3-brane stack.
Without the $Z$ factors, this spacetime is a solution if 
$\del_M\del^MN=0$, where $M=i,m$: these are essentially plane waves 
with inhomogeneities, and including the $Z$ factors gives D3-brane 
analogs thereof. Recalling that both $AdS_5\times S^5$ and plane 
wave spacetimes are $\al'$-exact string backgrounds, it is likely 
that the backgrounds (\ref{AdSnullLif}), and (\ref{AdSplanewave}) 
in the near horizon limit, are also $\al'$-exact. To elaborate on 
this, we note that the curvature invariants $R,\ R_{MN} R^{MN},\ 
R_{MNPQ} R^{MNPQ}$, for these null-deformed spacetimes are finite and
identical to those of $AdS_5\times S^5$. Any higher derivative
correction to the action is expected to stem from covariant
contractions involving the additional curvature components: however
since the metric itself (and also \eg\ the Ricci tensor etc) has no
nonzero component with multiple upper $+$-indices, these additional
contractions vanish, thus giving no further corrections to
$AdS_5\times S^5$ which itself is $\al'$-exact \cite{Kallosh:1998qs}. 
While this is not a proof that all higher derivative corrections to 
(\ref{AdSnullLif}) vanish, it is suggestive, making these $AdS$ 
plane waves potentially more interesting (some earlier work 
on similar $\al'$-exact backgrounds appears in \cite{orlando}).

We will mainly focus here on null deformations with $g_{++}(x_i,r)$, 
which can all be thought of as simply solutions to 
5-dim gravity with negative cosmological constant, satisfying\\ 
$R_{MN} = -4g_{MN}$\ $(M,N=\mu,r)$, the cosmological constant arising 
from the 5-form flux, the $X^5$ part effectively untouched. Thus 
they are all solutions to a 5-dim effective action\ 
$S_5\sim \int d^5x \sqrt{-g}\ (R + 12)$.
[More generally, we can consider $g_{++}(r,x_i,x^+,\Omega_l)$, sourced 
by other matter fields.]
The 5-dim part of the spacetimes (\ref{AdSnullLif}) are then 
solutions to $R_{MN} = -4g_{MN}$ if $g_{++}$ satisfies
\be\label{eomg++}
r^2 \del_r^2g_{++} + r \del_rg_{++} - 4g_{++} + r^2 \del_i^2 g_{++} = 0\ .
\ee
$\del_-, \del_+$ are Killing vectors if $g_{++}=g_{++}(r,x_i)$. With 
$g_{++}>0$, it is natural to take $x^-$ as the time direction\ 
($g^{--}=-r^4g_{++}<0$ implies that $x^-$-constant surfaces are 
spacelike), with $x^+$ then a spatial direction. The deformation 
being lightlike is special: for instance, it is noteworthy 
that the equation above is linear, although this is in the full 
nonlinear gravity theory (not just in a linearized approximation). 
It is also noteworthy that turning on the $g_{++}$ mode (keeping it
static in $x^-$-time) does not source any other metric component: 
this is a consistent closed subsystem in itself. 
All these solutions preserve some supersymmetry, as will be outlined 
in the next subsection: in particular, the inhomogenous solutions 
preserve the same amount of supersymmetry as the homogenous one, so 
in some sense there is a moduli space of solutions here (although 
in a reduced effective action, $g_{++}$ does not enter, being 
lightlike).

We want to focus on normalizable solutions for $g_{++}$, \ie\ 
$g_{++}\ra {1\over r^2} r^4 f_4(x_i)$ near the boundary $r=0$. 
Using the usual $AdS/CFT$ dictionary, these will then have the 
interpretation of states in the \Nf\ SYM theory. To elaborate on 
this, consider an asymptotically $AdS_5$ solution with metric of 
the form
\be\label{FeffGr}
ds^2={dr^2\over r^2} + h_{\mu\nu} dx^\mu dx^\nu 
= {dr^2\over r^2} + {1\over r^2} 
(g^{(0)}_{\mu\nu}+r^2g^{(2)}_{\mu\nu}+r^4g^{(4)}_{\mu\nu}+\ldots) 
dx^\mu dx^\nu\ ,
\ee
written in the Fefferman-Graham expansion about the boundary $r=0$. Then 
holographic renormalization methods \cite{Balasubramanian:1999re,
Myers:1999psa,Emparan:1999pm,de Haro:2000xn,Skenderis:2002wp} 
give relations between the metric coefficients $g^{(k)}_{\mu\nu}$
and the holographic stress tensor calculated as\ 
$T_{\mu\nu} \sim \lim_{r\ra 0}\ {1\over r^2} {1\over G_5} 
(K_{\mu\nu}-Kh_{\mu\nu}-3h_{\mu\nu})$, where $K_{\mu\nu}$ is the extrinsic 
curvature\footnote{\label{f1} We have\
$K_{\mu\nu}=-{1\over 2} (\nabla_\mu n_\nu + \nabla_\nu n_\mu)$,\ with
$n_\mu$ the outward pointing unit normal to the surface $r=const$. 
For the boundary being $r=0$, we have\ $n=-{dr\over r}$ : this gives\ 
$K_{\mu\nu}={r\over 2} h_{\mu\nu,r}$.}. For (\ref{AdSnullLif}), 
the only departures from the $AdS_5$ expressions are in 
$\{++\}$-components and we have
\be\label{T++}
K_{++}=r^2 f_4(x_i)\ \ \quad \Rightarrow\qquad 
T_{++} = {f_4(x_i)\over 4\pi G_5}\ .
\ee
This stress tensor, roughly encoding a ``chiral'' wave, is
automatically traceless and conserved.\ Thus these null deformed
spacetimes all correspond to waves on the boundary with nonzero
constant energy momentum component $T_{++}$, varying inhomogenously in
the $x_i$-plane. In other words, these correspond to lightcone states
in \Nf\ SYM theory, obtained by turning on finite lightcone momentum
density, possibly with inhomogeneities. The homogenous $AdS$ plane
wave solution studied in \cite{Narayan:2012hk} corresponds to uniform
lightcone momentum density. The total lightcone momentum is\ $P_+\sim
\int d^2x_i dx^+ T_{++}$. No other observable has an expectation value
in the dual field theory in these states. The fact that the
$g_{++}$-mode comprises a closed subsystem in itself appears to be a
reflection of the fact that lightcone momentum density (constant in
lightcone time) can be consistently turned on in lightcone SYM without
sourcing other operators. It is worth noting that other
$x^+$-dependent deformations of $AdS/CFT$ may well exist, with nonzero
energy-momentum component $T_{++}$, suggesting that the present
discussion is not a complete classification of lightcone states in
\Nf\ SYM: however these will generically have other energy-momentum
components nonzero too, thus differing from the class here.

One might expect the local lightcone momentum density in the field 
theory in a region $\Delta x_i$ to always be positive, since the local 
dispersion relation\
$p_-={p_i^2\over 2p_+}$ ,\ where $p_+\sim \int_{\Delta x_i} T_{++}$,\ 
would suggest negative energy states otherwise. 
Imposing this, we have 
\be
T_{++}\geq 0
\ee
everywhere in space: this also effectively follows from a null energy
condition on the boundary.  This is a nontrivial physical condition on
the solutions, as we see below. Another physical criterion is that the
energy-momentum density is bounded: this is also not generically true,
and imposing this restricts the family of solutions.

Towards understanding solutions, we note that the $g_{++}$-equation 
is linear: this means that various ``basis'' solutions can be 
superposed to give composite solutions. To find the basis solutions, 
consider a separable ansatz\ $g_{++}=h_{++}(r) f(x,y)$, which gives
\be\label{h++eom} 
r^2 h_{++}'' + r h_{++}' - (4+k^2r^2) h_{++} = 0\ ,\qquad
\del_i^2 f + k^2f = 0\ .
\ee
for some constant parameter $k$. For $k=0$ and $f=const$ (\ie\ no 
$x_i$-dependence), we have $g_{++}= h_{++}=Qr^2$ as the normalizable 
solution: this is the homogenous $AdS$ plane wave discussed in 
\cite{Narayan:2012hk}, with $T_{++}\geq 0$ implying $Q\geq 0$. The 
non-normalizable solution $g_{++}\sim {1\over r^2}$ is in fact 
simply $AdS$ after a coordinate transformation. Inhomogenous 
solutions also exist for $k=0$, as we will see below.

For nonzero $k$, the radial equation is a Bessel equation in general. 
A subset of basis modes is then obtained with $k^2=k_1^2+k_2^2>0$: 
this gives\ $f(x,y)=\sin k_1x \sin k_2y$, and $h_{++}(r)=K_2(kr), I_2(kr)$.
Restricting to normalizable solutions (noting the asymptotics\ 
$I_2(kr)\sim k^2r^2$ as $r\ra 0$) picks out modes of the form\ 
$g_{++}=I_2(kr) \sin k_1x \sin k_2y$\ (or $g_{++}=I_2(kr) e^{ik_1x+ik_2y}$).
Thus for given $(k_1,k_2)$, the $x_i$-plane is lattice-like in a 
sense, with a unit cell of size\ $\sim {1\over k_1}\times {1\over k_2}$.
The general solution of this sort, and its asymptotic form near the 
boundary $r=0$, is
\be\label{g++Fourier}
g_{++} = \int d^2k_i\ {\tilde f}(k_1,k_2) I_2(kr) e^{ik_1x+ik_2y}\ 
\ra_{r\ra 0}\ \ r^2 \int d^2k_i\ {\tilde f}(k_1,k_2) k^2 e^{ik_1x+ik_2y}\ ,
\ee
the $k^2 {\tilde f}(k_1,k_2)$ being dimensionful (complex) Fourier 
coefficients of the function $f_4(x_i)$ (using (\ref{T++})) over 
the plane wave components $e^{ik_ix_i}$:\ the homogenous solution 
is given by ${\tilde f}(k_1,k_2)={Q\over k^2} \delta^2(k_i)$. In 
general, the coefficients $k^2 {\tilde f}(k_1,k_2)$ have dimensions 
of (boundary) energy density.
A generic function $f_4(x_i)$ can be constructed from this basis, 
giving the most general lightcone momentum density $T_{++}\sim f_4(x_i)$\ 
(\ref{T++}). In other words, a general spatially varying $T_{++}$ 
maps to a (static) bulk dual metric with $g_{++}$ as above.

To put the current discussion in perspective with that on normalizable
modes in \eg\ \cite{Balasubramanian:1998sn}, consider a massless
scalar field $\varphi$ in $AdS_5$ in lightcone coordinates: then
${1\over\sqrt{-g}} \del_\mu (g^{\mu\nu}\sqrt{-g} \del_\nu\varphi)=0$
for modes\ $\varphi=e^{ik_+x^++ik_-x^-+ik_ix_i} R(r)$\ simplifies to
\be
R''-{3\over r} R' - k^2 R = 0\ , \qquad 
k^2\equiv k_\mu k^\mu = -2k_+k_-+k_i^2\ .
\ee
Redefining $R=r^2 y(r)$ casts this radial equation as\ 
$r^2y''+ry'-(4+k^2r^2)y=0$, \ie\ in the same form as that in 
(\ref{h++eom}).
Now with $k^2=0$, we have\ $y=r^2$ as the normalizable solution. 
With $k^2<0$, this gives $y=J_2(kr)$ as normalizable modes: however 
$k^2=-2k_+k_-+k_i^2<0$, which requires $k_\pm\neq 0$, so these are 
fluctuating modes. If we look for static, $x^-$-independent modes, 
then we have $k^2=k_i^2\geq 0$, and the radial equation gives 
$y=I_2(kr)$ as the normalizable solution for nonzero $k^2$, \ie\ 
$\varphi=r^2I_2(kr) e^{ik_ix_i}$\ (or $\varphi=r^4$ for $k^2=0$). 
These are thus normalizable backgrounds (rather than fluctuating
modes), which have the interpretation of time-independent states in
the dual gauge theory.  Generic metric perturbations (at linearized
order) are governed by a similar equation, so their behaviour for
normalizable background solutions is similar. The $g_{++}$-mode is
special, in that it is a closed subsystem in itself, with its
equations at linearized order (\ref{eomg++}), (\ref{h++eom}), in fact
being exact in the full nonlinear theory.

These normalizable background solutions to (\ref{h++eom}) grow
exponentially in the interior, and one might be concerned if these
inhomogenous solutions are physically allowed. In this light, we
recall that the homogenous $AdS$ plane wave with $g_{++}=Qr^2$ also
grows in the interior, although only as a power law. In general, from
above, we see that a normalizable static background will grow in the
interior. Also from the dual field theory point of view, we are
describing configurations with nontrivial lightcone momentum density
$T_{++}$ distinct from the vacuum, the homogenous $AdS$ plane wave
corresponding to uniform $T_{++}$. The general spatially varying
boundary $T_{++}$ which is bounded, conserved and satisfying sensible
energy conditions and thus an allowed configuration in field theory,
maps quite generally as (\ref{g++Fourier}) above to these static 
normalizable backgrounds using these basis modes of (\ref{h++eom}),
which are the only static normalizable solutions with $k^2>0$ 
corresponding to oscillatory $e^{ik_ix_i}$-type boundary spatial 
behaviour. Small (linearized) fluctuations about $AdS_5$ with only 
$g_{++}$ nonzero (\ie\ only $T_{++}$ nonzero) of the form 
$e^{ik_+x^++ik_-x^-+ik_ix_i} h_{++}(r)$ including lightcone time $x^-$
dependence are also governed by (\ref{h++eom}) at linear order, the
other equations forcing $k_-=0$. This could mean that small 
$g_{++}$-fluctuations also source other modes at linear order: 
however it also implies that static $g_{++}$-backgrounds form a 
closed subsystem.
For $k^2=k_i^2<0$ in (\ref{h++eom}), $f(x_i)$ contains hyperbolic
sinh/cosh-functions, the radial equation giving Bessel functions
$J_2(kr), Y_2(kr)$. Then basis modes\ $J_2(kr) \cosh k_1x \cosh k_2y$\
are normalizable at the boundary. Also for $k^2=0$, we have modes\
$r^2 (x^2-y^2)$\ or\ $r^2 \sin \chi x \sinh \chi y$ (for any $\chi$) 
as inhomogenous normalizable backgrounds. However, in these cases, 
the boundary lightcone momentum density $T_{++}$ for these basis 
modes is not bounded, growing indefinitely in certain regions for 
large $x,y$, which makes their status less clear.

From the asymptotics\
$I_2(kr)\sim {e^{kr}\over \sqrt{r}}\ \ (r\ra\infty)$, we have seen 
that these basis solutions always grow large in the interior 
$r\ra\infty$: thus $g_{++}$ is not always positive, since the sines 
oscillate and $I_2$ grows for large $r$. Thus there exist loci 
where $g_{++}=0$, quite generically: at these, $g^{--}= -r^4 g_{++}=0$,
so that constant-$x^-$ hypersurfaces which are spacelike for
$g_{++}>0$ become null. Alternatively, the vector $\del_-$ which is
timelike for $g_{++}>0$ becomes null at the loci $g_{++}=0$. These
loci are thus akin to $horizons$, where both $x^{\pm}$ are lightlike
directions. The vicinity of a horizon does not have anything unusual
happening, the spacetime becoming simply $AdS_5$ in lightcone
coordinates: since curvature invariants are finite for the entire 
spacetime, they are in particular finite in the vicinity of the 
horizon too (it would be interesting to study geodesics and tidal 
forces in detail here). Generic particle trajectories will cross 
the horizon: on crossing a horizon, $x^+$ becomes the natural time 
direction.
Consider a bulk particle trajectory with action\ 
$S={1\over 2} m \int d\tau\ g_{\mu\nu}{\dot x^\mu}{\dot x^\nu} = 
{1\over 2} m \int d\tau\ (2g_{+-}{\dot x^+}{\dot x^-}+g_{++}({\dot x^+})^2
+g_{IJ} {\dot x^I} {\dot x^J})$. If we fix lightcone gauge $x^+=\tau$ 
as often defined, we have the conjugate momenta\ $p_I=mg_{IJ}{\dot x^J}\ ,
\ p_-={\del L\over\del {\dot x^-}}=mg_{+-}$\ ($p_-<0$ if $g_{+-}<0$), 
and the Hamiltonian\ \bc
$H=p_-{\dot x^-}+p_I{\dot x^I}-L = {1\over 2m} g^{IJ} p_Ip_J - 
{1\over 2} mg_{++} = {1\over 2|p_-|} p_I^2 - {|p_-|\over 2} r^2g_{++}$ ,\ec
using $g_{II}=-g_{+-}={1\over r^2}$.\ We see that $g_{++}<0$ gives a 
positive potential while $g_{++}<0$ makes the potential term negative: 
the latter is in some sense an artifact of choosing lightcone 
gauge here fixing the spatial direction $x^+$ as time. Approaching 
the horizon locus, we obtain a free particle Hamiltonian, with the 
dispersion relation\ $H\equiv p_+\sim {1\over 2|p_-|} p_I^2$. 
Alternatively, fixing $\tau=x^-$, we have\ 
$S={1\over 2} m \int d\tau\ (2g_{+-}{\dot x^+}+g_{++}({\dot x^+})^2
+g_{IJ} {\dot x^I} {\dot x^J})$, and the conjugate momenta\ 
$p_I=mg_{IJ} {\dot x^J}\ ,\ 
p_+={\del L\over\del {\dot x^+}}=mg_{+-}+mg_{++}{\dot x^+}$ .
This gives the Hamiltonian\ $H=p_+{\dot x^+}+p_I{\dot x^I}-L = 
{1\over 2m} g^{IJ} p_Ip_J + {1\over 2mg_{++}} (p_+-mg_{+-})^2$. 
For $g_{++}\ra 0$ however, the last term arising from 
$g_{++} ({\dot x^+})^2$ disappears since now $p_+\ra mg_{+-}$, and 
we again recover the nonrelativistic dispersion relation\ 
$H\equiv p_-={1\over 2p_+} p_I^2$ near the $g_{++}=0$ horizon, as 
for $AdS_5$ in lightcone coordinates.

For any such $AdS$ plane wave, there are typically multiple such 
(disconnected) horizons, since the loci $g_{++}=0$ have many 
solutions. To illustrate this, consider for simplicity the solution
\be\label{Qsinkx}
g_{++} = Qr^2 + {\tilde f}_k I_2(kr) \sin kx\ ,
\ee
which has $y$-translations, but is $x$-striped. This is a 
superposition of a basis solution above and the homogenous $AdS$ 
plane wave discussed in \cite{Narayan:2012hk}. Near the boundary 
$r\ra 0$, we have\ $g_{++}\ra r^2 (Q + k^2{\tilde f}_k \sin kx)$, so 
that the holographic lightcone momentum density\
$T_{++}\sim Q + k^2 {\tilde f}_k \sin kx$, is positive for 
$Q> k^2{\tilde f}_k$.\ However since $I_2(kr)$ grows, a horizon 
$g_{++}=0$ develops in the interior: \eg\ within the unit cell 
$-{\pi\over k} \leq x\leq {\pi\over k}$ , the subregion 
$-{\pi\over k} \leq x\leq 0$ gives rise to a horizon locus 
$g_{++}=0$. In the solution above, this is
\be\label{QsinkxHor}
\sin kx = -{Qr^2\over {\tilde f}_k I_2(kr)}\ .
\ee
So we see that $x\ra 0$ as $r\ra\infty$ and $x\ra -{\pi\over 2k}$ 
as $r$ approaches the value where the r.h.s. becomes unity. In the 
$(r,x)$-plane, these are roughly half-ellipse-shaped curves, one 
in the appropriate subregion in each unit cell. In the subregion 
$0\leq x\leq {\pi\over k}$, the sines are positive and $g_{++}>0$ 
so there is no horizon.
For spacetimes with $g_{++}$ having both $x,y$-dependence, the 
horizon loci are given by the surface $g_{++}(r,x_i)=0$, or 
$r_0=r(x_i)$ as the implicit solutions.



Consider a spacetime of the form (\ref{AdSnullLif}), with a leading 
homogenous plane wave piece superposed with a ``small'' inhomogeneity 
in the $x$-direction, thinking of the inhomogeneity as a static 
perturbation ($y$-translation symmetry exists). 
We would like to define this by requiring that the energy-momentum 
or lightcone momentum density $T_{++}$ is only a small departure 
from constant density $Q$, \eg\ \
$T_{++} \sim\ Q + \epsilon f(x)$.   
When the parameter $\epsilon=0$, this is just the homogenous 
$AdS$ plane wave.  
Now since $T_{++}$ is essentially the asymptotic form of $g_{++}$, we 
effectively see that $g_{++}$ can be found using the Fourier 
coefficients of $f(x)$, using (\ref{g++Fourier}).
Consider thus Fourier modes of the form\ $A e^{-k^2 \sigma^2}$ :\ these 
could reflect a small (Gaussian) lump localized around $x=0$ with 
width $\sigma$, and thus a small perturbation for small amplitude 
$A$. We then see that\ 
\be\label{g++Smallxdep}
g_{++}\sim\ Qr^2 + A \int_{-\infty}^\infty dk\ e^{-k^2\sigma^2} 
I_2(kr) \sin kx \ra_{r\ra\infty} \ \ 
Qr^2 + A \int_{-\infty}^\infty dk\ e^{-k^2\sigma^2} {e^{kr}\over\sqrt{r}} 
\sin kx \ .
\ee
For $x\ra 0^+$, each individual component in the second term is 
positive and $g_{++}>0$. However evaluating this at $x\ra 0^-$ 
and approximating,\ we see that the sines are negative and the 
second term grows large and negative: thus we expect that 
$g_{++}=0$ somewhere. We see that $x\ra 0^{\pm}$ have substantially
different behaviour for a small inhomogenous perturbation about $x=0$
near the boundary, with in fact large effects in the interior. This
suggests that in fact the homogenous $AdS$ plane wave is special:
apparently ``nearby'' $AdS$ plane waves with small inhomogenous
modifications are in fact large departures with qualitative
differences in the interior, such as the emergence of $g_{++}=0$
horizon loci.

It is also interesting to imagine a general near-boundary
Fefferman-Graham expansion
\be
g_{++} = {1\over r^2} (r^4 f_4(x_i) + r^6 f_6(x_i) + \ldots)
\ee
with the equation of motion (\ref{eomg++}) giving relations between
the coefficients $f_n(x_i)$,
\be
12 f_6 + \del_i^2 f_4 = 0\ ,\quad \ldots\ \quad 
n(n-4) f_n + \del_i^2 f_{n-2} = 0\ .
\ee
Then (with $y$-translations retained) there exist truncated 
solutions of the form\ $g_{++}=r^2f_4(x)+r^4f_6(x)$,\ where\
$f_6''=0 ,\ f_4''+12f_6=0$: an example is\ $g_{++}=Q(6r^2x^2-r^4)$. 
This has\ $T_{++}\sim Qx^2$, and a $g_{++}=0$ horizon locus\
$x^2={1\over 6} r^2$, which intersects the boundary at $r=0$, and 
so appears somewhat different from \eg\ (\ref{QsinkxHor}) which 
has support only in the interior.




Our discussion so far has been primarily for $AdS_5$ null
deformations. $AdS_D$ null deformations of the form (\ref{AdSnullLif})
are solutions if
\be\label{eomg++D}
r^2 \del_r^2g_{++} + (6-D) r \del_rg_{++} - (2D-6) g_{++} 
+ r^2 \del_i^2 g_{++} = 0\ .
\ee
Modes with spatial dependence $e^{ik_ix_i}$ give a radial equation\ 
$r^2h_{++}''+(6-D)rh_{++}'-(2D-6+r^2k^2)h_{++}=0$. The normalizable 
background solutions are then\ $r^{(D-5)/2} I_{{D-1\over 2}}(kr) e^{ik_ix_i}$. 
Near the boundary, these asymptote to\ $r^{D-3}$, while for large 
$r$, we obtain exponential growth\ $e^{kr}$.
For $D=3$, there are no spatial directions $x_i$, and the equation 
above reduces to\ $r^2g_{++}''+3rg_{++}'=0$, which gives\ $g_{++}=const$ 
as the normalizable solution, which is automatically homogenous 
(aspects of null or chiral deformations of conformal field theories 
appear in \eg\ \cite{Costa:2010cn}).
In general, these null deformations amount to turning on nontrivial
expectation values for lightcone momentum density $T_{++}$ in the dual
conformal field theory:\ for M2- and M5-branes, these $AdS_4\times
X^7$ and $AdS_7\times X^4$ plane waves would seem to correspond to
states with lightcone momentum density $T_{++}$ in the Chern-Simons
(ABJM-like) and the $(2,0)$ theories respectively. It would be 
interesting to explore these further.

We have been discussing normalizable backgrounds with broken 
translation invariance: in this context, it is worth noting 
\cite{Iizuka:2012iv} which discusses homogenous but anisotropic 
backgrounds which arise from extremal branes with a Bianchi 
classification.



\subsection{Supersymmetry}

We want to study supersymmetry properties of the 10-dim spacetime 
(\ref{AdSplanewave}). Defining vielbeins\
\be
e^+=Z^{-1/4} dx^+ ,\quad e^-=Z^{-1/4} (dx^--{N\over 2} dx^+) ,\quad
e^i=Z^{-1/4} dx^i ,\quad e^m=Z^{1/4} dx^m\ ,
\ee
which are natural for the lightcone coordinates here. (For 
$N(x^i,x^m)=0$, this is simply the D3-brane stack solution in 
lightcone coordinates.)\ Using\ $de^a=-\omega^a{_b}\wedge e^b$,\ 
this then gives the spin connection\
\bea
&& \qquad \omega_{-m} = {1\over 4} Z^{-1/4} \del_m\log Z e^+\ , \quad 
\omega_{im} = -{1\over 4} Z^{-1/4} \del_m\log Z e^i\ , \nonumber\\
&&\qquad\qquad\qquad
\omega_{mn} = {1\over 4} Z^{-1/4} (\del_n\log Z e^m - \del_m\log Z e^n)\ ,
\nonumber\\
&& \omega_{+m} = {1\over 4} Z^{-1/4} \del_m\log Z e^- 
+ {1\over 2} Z^{-1/4} \del_m\log N e^+\ , \quad 
\omega_{+i} = {1\over 2} Z^{1/4} \del_i\log N e^+\ , \quad
\eea
We only need to consider gravitino variations given the nontrivial 
fields in these backgrounds. Then 
$\delta \psi_M = {1\over\kappa} D_M\epsilon + {i\over 480} \gamma^{M_1\ldots M_5}
F_{M_1\ldots M_5} \gamma_M \epsilon = 0$, shows that the variations\
$\delta\psi_-, \delta\psi_i, \delta\psi_m$ are the same as for the 
case $N=0$. Evaluating $\delta\psi_+$ gives new terms containing\
$(\omega_{+m+}\Gamma^{+m}+\omega_{+i+}\Gamma^{+i})\epsilon$ in addition to
the terms for $N=0$. All these conditions can thus be satisfied if 
the spinor satisfies\ $\Gamma^4\epsilon=\epsilon$\ (as for the usual 
D3-brane solution), and\ $\Gamma^+\epsilon = 0$. The latter is the 
familiar spinor condition for null solutions. What we thus see is 
that the inhomogenous $AdS$ plane waves preserve just as much 
supersymmetry as the homogenous one\footnote{It is worth noting 
constructions such as \cite{Itsios:2012dc} which may be useful 
in generating new solutions of this sort, perhaps with reduced 
supersymmetry.}.

\section{On $x^+$-dimensional reduction}

The generic 10-dim solution (\ref{AdSnullLif}) can be dimensionally 
reduced on $X^5$ to give a 5-dim system: with $g_{++}(r,x_i)$ having 
no $x^+$-dependence, this is a solution to 5-dim gravity with negative 
cosmological constant (and no other matter sources). Now we consider 
regarding $x^+$ as a compact direction and dimensionally reducing\ 
$\int d^5x \sqrt{-g^{(5)}}\ (R^{(5)} - 2\Lambda)$\ on it as\
\bc
$\int dx^+ d^4x \sqrt{-g^{(4)}}\ (R^{(4)} - \#\Lambda e^{-\phi} 
- \# (\del\phi)^2 - \# e^{3\phi} F_{\mu\nu}^2 ) ,$
\ec
where the 4-dim metric undergoes a Weyl transformation as\ 
$g^{[4]}_{\mu\nu}=e^{\phi} g^{[5]}_{\mu\nu}$\ (and the numerical 
constants $\# $ can be fixed). This gives the resulting 4-dim 
Einstein metric (relabelling $x^-\equiv t$) as
\be\label{4dmet}
ds_4^2 = \sqrt{g_{++}} \left(-{dt^2\over r^4 g_{++}} + {dx_i^2+dr^2\over r^2} 
\right) ,\qquad\ \ 
e^\phi = \sqrt{g_{++}}\ ,\qquad A_t = - {dt\over r^2 g_{++}}\ ,
\ee
with $g_{++}(r,x_i)$, and the overall conformal factor $\sqrt{g_{++}}$ 
in the metric arises from the KK-scalar $e^\phi$. 
For the homogenous case with $g_{++}=Qr^2$, \ie\ the $AdS$ plane wave,
discussed in \cite{Narayan:2012hk}, this gives the hyperscaling
violating metric lying in the family $\theta=d-1$.  The KK scalar
grows in the interior as\ $\phi\sim \log r$ in this case.  All
normalizable $g_{++}$ solutions have $g_{++}\ra_{r\ra 0} r^2f_4(x_i)$
so their radial dependence near the boundary is similar to the
homogenous $AdS$ plane wave case: near $r\ra 0$, the lower 
dimensional description (\ref{4dmet}) approaches
\be
ds_4^2 = -{dt^2\over r^5 \sqrt{f_4(x_i)}} + {\sqrt{f_4(x_i)}\over r} 
(dx_i^2+dr^2)\ ,\quad \ \
e^\phi = r \sqrt{f_4(x_i)}\ ,\quad A_t = - {dt\over r^4 f_4(x_i)}\ .
\ee
These static solutions from the lower dimensional point of view are of
course of a specific kind (\eg\ gauge field being solely electric),
with some function $g_{++}(r,x_i)$ introducing inhomogeneities.
Presumably more general inhomogenous backgrounds can be found in the 
lower dimensional theory: their uplift might not fit into this $AdS$ 
null deformation pattern upstairs of course.

The more general inhomogenous cases discussed here give rise to 
inhomogenous solutions after dimensional reduction too. However in these 
cases, several issues arise generically, the most basic one being that 
$g_{++}\ra 0$ somewhere in the interior implying that the $x^+$-circle 
shrinks. A little investigation suggests that generically, the surface 
$g_{++}=0$ (upstairs) does not give rise to a smooth cigar-like geometry 
in the $(r,x^+)$-directions: generically $g_{++}=0$ does not also coincide 
with $\del_rg_{++}=0$. For concreteness, assuming a Fefferman-Graham-like 
expansion\ \ $g_{++} \sim (r-r_0)^2 f_2 + \ldots$,\ for 
$g_{++}$ exists near $r=r_0(x_i)$ where $g_{++}(r_0)=0$, the equation 
of motion (\ref{eomg++}) is not satisfied near $r=r_0$ unless $f_2=0$.
These seem worse than conical singularities in the interior, with 
$g_{++}\sim c(r-r_0)+\ldots$ in the vicinity of the horizon $g_{++}=0$.
This is not in contradiction with the fact that all curvature
invariants vanish everywhere (upstairs), which is a consequence of 
the lightlike nature of this entire class of solutions. Thus these
configurations generically do not seem like analogs of the 
$AdS$-soliton where the geometry caps off smoothly at some $r=r_0$. 
However this is worth exploring further.

For $x^+$-compact, we have a shrinking circle, in which vicinity\ 
$g_{++}\sim c(r-r_0)+\ldots$: this means that in the
vicinity of this location, string modes winding around the $x^+$-circle 
become light so that the gravity solution cannot be trusted. From the 
dual field theory perspective, various operators dual to these string 
states acquire low anomalous dimensions so do not decouple from the 
effective low energy dynamics. It would seem that these new light 
states would significantly alter the system. Assuming that the 
effects of these light states are approximately localized and do not 
destabilize the entire spacetime, one might imagine that the metric 
in (\ref{4dmet}) is approximately valid away from the locations where 
the circle pinches off. In general, this gives inhomogenous 
domain-like structures in the $x_i$-plane. For concreteness, consider 
the 5-dim spacetime in (\ref{Qsinkx})\ (ignoring the $X^5$), 
which has horizons as $g_{++}\ra 0$ in the spatial domains where 
$\sin kx<0$. Then near the horizon, we have\
$g_{++}\sim c (r-r_0) + \ldots$, where $r_0(x_i)$ in general varies 
as a function of the spatial location $x_i$: from (\ref{4dmet}), 
we then have
\be
ds_4^2 \sim \sqrt{r-r_0} \left(-{dt^2\over r^4 (r-r_0)} + 
{dx_i^2+dr^2\over r^2} \right) ,\quad\
e^\phi \sim \sqrt{r-r_0}\ ,\quad\ A_t \sim - {dt\over r^2 (r-r_0)}\ .
\ee
Similarly, in the spatial domains where $g_{++}>0$ everywhere, we 
have\ $g_{++}\ra_{r\ra\infty} {e^{kr}\over\sqrt{r}} \sin kx$: then the 
asymptotic lower dimensional metric and KK-scalar from (\ref{4dmet}) are
\be
ds_4^2 \ra_{r\ra\infty} - {dt^2 \over r^{9/2} e^{kr/2} \sqrt{\sin kx}}
+ {e^{kr/2}\over r^{5/2}} \sqrt{\sin kx} (dx_i^2+dr^2)\ ,\quad
e^\phi \sim {e^{kr/2}\over r^{1/4}} \sqrt{\sin kx}\ .
\ee
There is in general a lattice-like structure in the $x_i$-plane 
for $g_{++}(r,x_i)$ having periodic structure in the $x_i$: 
the locations where $g_{++}=0$ define unit cell boundaries in a sense.
Near these boundaries, we have $g_{tt}\ra{1\over\sqrt{g_{++}}}$ while 
$g_{ii}, g_{rr}\ra \sqrt{g_{++}}\ra 0$, implying that the spatial 
directions shrink, while the time direction grows, \ie\ lightcones 
open up.


\subsection{Asymptotically Lifshitz solutions with inhomogeneities}

Consider now $g_{++}(x^+,x_i,r)$ to be a function of $x^+$ also: then 
the equation for $g_{++}$ continues to be (\ref{eomg++}), which most 
notably has no $x^+$-derivatives. This means one can separate variables 
here, implying that $AdS$ deformations with metric of the form 
(\ref{AdSnullLif}), and
\be\label{g++lif}
g_{++}=F(x^+) + g_{++}(x_i,r)\ ,
\ee
are all solutions, with the non-normalizable deformation 
$F(x^+)$ being sourced through $R_{++}$ by various lightlike 
sources contributing to $T^{bulk}_{++}$. 
The leading term $F(x^+)$ (which could be an $x^+$-independent
constant as well) by itself gives rise, after $x^+$-dimensional
reduction, to $z=2$ 4-dim Lifshitz spacetimes as discussed in the
string constructions \cite{Balasubramanian:2010uk,Donos:2010tu}
holographically dual to the corresponding deformations (and DLCQ) of
SYM theories. Then, if we restrict to normalizable solutions in the
second term $g_{++}(x_i,r)$ as in the previous sections, we have\
$g_{++}\sim_{r\ra 0} F(x^+) + r^2 f_4(x_i)$: in this case, these
solutions after $x^+$-dimensional reduction, are all asymptotically
Lifshitz. Thus they are best thought of as states in the 
Lifshitz-like (non-normalizable) null-deformations (and DLCQ) of 
the SYM theory.

We now first describe some aspects of these Lifshitz string
constructions involving $AdS$ null deformations, focussing therefore
only on the leading non-normalizable term $F(x^+)$ in $g_{++}$: this
gives some $AdS/CFT$ perspective on the Lifshitz vacuum in these
constructions. First we note that this term $F(x^+)$ does not
contribute at all to the holographic stress tensor in these
backgrounds, giving $T_{++}=0$ (including counterterms for the
sources): \ie\ the holographic stress tensor vanishes despite the fact
that there are nontrivial lightlike matter fields (\eg\ a null
dilaton) sourcing the null deformation. To elaborate, consider first
the case \cite{Balasubramanian:2010uk} where $F(x^+)$ is sourced by a
lightlike scalar (\eg\ the dilaton): then the stress tensor, using
\cite{Awad:2007fj}, is\
\be\label{lifTmunu}
T_{\mu\nu} \sim\ {1\over  G_5}
\left(K_{\mu\nu} - K h_{\mu\nu} - 3 h_{\mu\nu} 
+ {1\over 2} G_{\mu\nu} - {1\over 4}\del_\mu\Phi\del_\nu\Phi 
+ {1\over 8} h_{\mu\nu} (\del\Phi)^2\right) ,
\ee 
where $G_{\mu\nu}, \del$ here are defined w.r.t. the boundary metric 
$h_{\mu\nu}$, and the scalar terms arise from counterterms involving 
the scalar (and $K_{\mu\nu}$ is the extrinsic curvature as before, see 
footnote~\ref{f1})\footnote{See \eg\ \cite{Ross:2011gu,Braviner:2011kz,
Baggio:2011cp,Chemissany:2012du} for recent discussions of holographic 
renormalization in Lifshitz backgrounds; see also \cite{Costa:2010cn}.}. 
The stress tensor can then be seen to vanish if, in (\ref{g++lif}), 
the leading term $F(x^+)$ alone is nonvanishing.

Alternatively, using the coordinate transformation\ 
$r=w e^{-f/2} , \ x^-=y^--{r^2 f'\over 4}$ , the metric (\ref{AdSnullLif}) 
with \eg\ $g_{++}={1\over 4} (\del_+\Phi)^2$ can be recast as\
\be\label{metconf5}
ds^2={1\over w^2} [e^{f(x^+)} (-2dx^+dy^- + dx_i^2) + dw^2] + d\Omega_5^2, 
\qquad\ \Phi=\Phi(x^+)\ ,
\ee
with the constraint 
${1\over 2}(\del_+f)^2-\del_+^2f={1\over 2}(\del_+\Phi)^2$. 
In this form, comparing with the Fefferman-Graham expansion 
(\ref{FeffGr}), the metric can be seen to have\ 
$g^0_{\mu\nu}=e^{f(x^+)}\eta_{\mu\nu}$ with all subleading coefficients 
identically vanishing: this solution is part of the general family of 
solutions
\be\label{geomscal}
ds^2={1\over r^2} ({\tilde g}_{\mu\nu}(x^\mu) dx^\mu dx^\nu + dr^2) 
+ d\Omega_5^2\ , \qquad \Phi=\Phi(x^\mu)\ ,
\ee
(after relabelling $w$ as $r$) with the $X^5$ and 5-form suppressed, and 
the conditions\ ${\tilde R}_{\mu\nu}={1\over 2} \del_\mu\Phi\del_\nu\Phi,\ 
\Box\Phi=0$\ \cite{Das:2006dz,Awad:2007fj} arising from the IIB supergravity 
equations of motion. These conditions can also be obtained using 
holographic renormalization methods \cite{de Haro:2000xn,Skenderis:2002wp}
and requiring that all subleading coefficients vanish from the 
Fefferman-Graham expansion for both metric (\ref{FeffGr}) and scalar\ 
$\Phi=r^{(d-\Delta)/2} (\Phi^0+r^2\Phi^2+\ldots)$, 
by solving\ $R_{MN}=-4g_{MN}+{1\over 2}\del_M\phi\del_N\phi$\ 
iteratively: this gives
\be\label{g2constr}
g_{\mu\nu}^2\sim R_{\mu\nu}^0-{1\over 2}\del_\mu\Phi\del_\nu\Phi
-{1\over 2(d-1)}\Big(R-{1\over 2} (\del\Phi)^2\Big) g_{\mu\nu}^0 :\qquad 
g^2_{\mu\nu}=0\ \ \Rightarrow\ \ 
R_{\mu\nu}^0={1\over 2}\del_\mu\Phi\del_\nu\Phi\ ,
\ee
 (for a massless scalar $\Delta=d$) also implying the higher order 
coefficients vanish if $g^{(4)}=0$. 
Likewise, with $\Box^0$ being the Laplacian w.r.t. $g^0_{\mu\nu}$, 
we also obtain\ $\Phi^{(2)}\sim \Box^0\Phi^0$: thus $\Phi^{(2)}=0$ 
implies $\Box^0\Phi^0=0$.\ 
In these coordinates, the boundary metric is\
$h_{\mu\nu}={1\over r^2} {\tilde g}_{\mu\nu}$, we see that the extrinsic
curvature is\ $K_{\mu\nu}=-h_{\mu\nu}$, so that the first three terms 
in $T_{\mu\nu}$ given in (\ref{lifTmunu}) cancel identically while 
the the last three terms also cancel using the constraint relation\ 
${\tilde R}_{\mu\nu}={1\over 2} \del_\mu\Phi^{(0)}\del_\nu\Phi^{(0)}$. 
Thus the stress tensor vanishes identically for these leading
non-normalizable deformations $g_{++}=F$.

The solutions (\ref{metconf5}), (\ref{geomscal}), thus appear
constrained from this point of view, with only the first coefficient
$g^{(0)},\ \Phi^{(0)}$ nonzero for all $r$, the subleading pieces of
the metric and scalar vanishing. These conditions on the
$g_{\mu\nu}^n,\ \Phi^n,\ n>0,$ are non-generic and appear to be
nontrivial constraints fine-tuning the CFT state after turning on the
sources $g^{(0)},\ \Phi^{(0)}$. These arguments have also been
discussed in \cite{Narayan:2012dh} in the cosmological singularities
context of \cite{Das:2006dz,Awad:2007fj}.

This discussion on the Fefferman-Graham expansion is mainly based on
the foliation where we have $g^{(0)}\neq 0$ alone: however the
relation between $g^{(2)}, \Phi^{(0)}$ can be seen to be true even
otherwise. Although we have presented these arguments for the case
where $F(x^+)$ is sourced by a lightlike dilaton, it appears likely
that similar arguments can be made for other matter sources discussed
in \cite{Donos:2010tu}, using appropriate matter counterterms to
construct the holographic stress tensor free of ultraviolet
divergences.  From this point of view, the Lifshitz vacuum has been
obtained by deforming using lightlike sources and also setting the
stress tensor to vanish (by requiring vanishing of \eg\
$g^{(4)}_{\mu\nu}$), thereby tweaking the state too.

It is worth making some general comments here. As discussed in
\cite{Awad:2007fj}, the coordinate transformations above, recasting
the metric to have a conformally flat boundary, are
Penrose-Brown-Henneaux (PBH) transformations
\cite{Penrose:1986ca,Brown:1986nw,Imbimbo:1999bj,skenderis,Fukuma:2002sb}:
these are bulk diffeomorphisms acting as a Weyl transformation on the
boundary.  In general, a PBH transformation changes the bulk foliation
and thus the near-boundary Fefferman-Graham expansion. In particular,
a spacetime where the boundary metric $g^{(0)}_{\mu\nu}$ is
conformally flat, as in (\ref{metconf5}), (\ref{geomscal}), will
generically have nonzero subleading coefficients $g^{(n)}_{\mu\nu} ,\
n>0$, after a PBH transformation: in general, this will lead to a
nonvanishing stress tensor as well, as studied in some of the
time-dependent situations in \cite{Awad:2007fj}.  From this point of
view, the stress tensor continuing to vanish for the null deformations
here is in some sense accidental, stemming from the fact that the PBH
transformation here results in a finite series, the metric in the PBH
coordinates (\ref{AdSnullLif}) truncating at order $g^{(2)}_{++}$.
Therefore the above arguments on constraining the state appear
foliation-dependent. Perhaps it is fair to say that the existence of a
coordinate choice, or foliation, where the Fefferman-Graham expansion
appears constrained (\eg\ with the corresponding holographic stress
tensor vanishing) is not generic. It would be worth understanding
these issues better.

This appears to have some bearing on the Lifshitz singularity due to
diverging tidal forces as $r\ra\infty$ \cite{Horowitz:2011gh}. In the
conformal coordinates (\ref{geomscal}), we see that the $AdS$
deformation with $g^{(0)}_{\mu\nu}={\tilde g}_{\mu\nu}$ alone could 
potentially lead to singularities on the Poincare horizon 
$r\ra\infty$, with the curvature invariants\ 
$R\sim\ r^2 {\tilde g}^{\mu\nu}\del_\mu\Phi\del_\nu\Phi + O(r^0)$,\ \ 
$R_{ABCD} R^{ABCD} \sim\ r^4 {\tilde R}_{\mu\nu\al\beta} {\tilde
R}^{\mu\nu\al\beta} + O(r^0)$ etc diverging\ (this can be seen by 
expanding out the curvature components and using the equation
${\tilde R}_{\mu\nu}={1\over 2} \del_\mu\Phi\del_\nu\Phi$:\ see also 
\cite{Narayan:2012dh} in the cosmological singularities context 
\cite{Das:2006dz,Awad:2007fj}). Equivalently, a metric that
is regular everywhere, with an expansion of this form about $r=0$,
must have the coefficients $g_{\mu\nu}^{(n)}$ generically nonzero if a
singularity is to be avoided at large $r$. For the null metric
(\ref{metconf5}), these curvature invariants 
vanish, since the lightlike solutions admit no nonzero
contraction. However this is a common feature of lightlike solutions,
as we may recall from plane-wave spacetimes. The singularities then
arise from diverging tidal forces. In this form, we see that the
divergence is due to the fact that the metric is constrained with all
subleading coefficients vanishing.  However it is worth noting that
the state in the $AdS/CFT$ perspective appearing constrained is not 
automatically equivalent to the string background being singular. 
Since the Lifshitz vacuum has been constructed by turning on
nontrivial sources, it would appear that strings would experience
nontrivial scattering due to the ``stuff'' making up the sources,
rendering the metric contribution alone likely incomplete: this has
been discussed in \cite{Bao:2012yt} in the context of the Lifshitz
string constructions in \cite{Hartnoll:2009ns}. In this light, the
Lifshitz backgrounds above are also likely to be good string
backgrounds, even if they appear constrained from an $AdS/CFT$ point
of view at the level of supergravity (see also \cite{Harrison:2012vy}
for other related discussions on Lifshitz singularities). For
instance, in the case \cite{Balasubramanian:2010uk} where $F(x^+)\sim
(\del_+\Phi)^2$ is sourced by a lightlike dilaton $\Phi(x^+)$ arising
from the NS-NS sector alone, the source would appear to be a
condensate of background strings making up the dilaton profile, while
the more general sources in \cite{Donos:2010tu} include R-R
backgrounds. It would be interesting to understand string propagation
in these backgrounds better. Relatedly it would also be interesting to
obtain a deeper understanding of the rules of $AdS/CFT$ and its
deformations/states for such apparently constrained backgrounds.

An interesting question in this regard pertains to the existence of
inhomogenous phases in asymptotically Lifshitz backgrounds, 
with broken spatial translation invariance.
In the context of $AdS$ null deformations, this can be simulated by 
$g_{++}$ modes of the form (\ref{g++lif}), which are in the full 
theory (not just in linearized gravity).
Once we turn on the subleading normalizable modes in (\ref{g++lif}), 
the stress tensor acquires a nonzero expectation value\ 
$T_{++}\sim {f_4(x_i)\over G_5}$, where $g_{++}(r,x_i)=F+r^2f_4+\ldots$.
These solutions have the same feature as before, \ie\ $g_{++}=0$ 
somewhere in the interior. After the $x^+$-dimensional reduction, 
this means that there are light string winding states that arise 
in the vicinity of these loci, which therefore suggest new stringy 
physics that modifies the gravity solution. It would be interesting 
to explore these further.

Besides null deformations (\ref{g++lif}), it is of interest to look
for different kinds of inhomogenous backgrounds, in particular say
striped phases. In this regard, note that technically, a 4-dim
Lifshitz spacetime is a solution to 4-dim gravity with negative
cosmological constant coupled to a massive vector: this is similar to
the field content for \eg\ charged 4-dim $AdS$ black brane solutions,
so one might imagine technical similarities between the two systems in
the analysis of linearized gravity perturbations, discussed for the
latter in \cite{Donos:2011bh}. In the Lifshitz context, the massive
gauge field and metric lift in the 5-dim metric-dilaton system to
metric modes, and we can then look for perturbations restricting to
simply normalizable metric modes ``upstairs'' (the scalar takes on
simply its non-normalizable background profile). Then for instance, it
can be seen that the modes \{$g_{+y}(x^+,x,r), \ g_{-y}(x,r)$\} give
rise to a closed subsystem of equations at the linearised level. These
map in the lower dimensional description to $\{ A_y, g_{ty}\}$.\
Consider an explicit ansatz for striped phases of the form (with
$F(x^+)={1\over 4} (\del_+\Phi)^2$ as in the leading solution),\ 
$g_{+y} = {1\over r^2} F h_{+y}(r) \sin kx ,\
g_{-y} = {1\over r^2} h_{ty}(r) \sin kx$,\
that simulate translation-invariance breaking states: this 
mimics possible regular clumps along the $x$-direction while 
translation invariance continues to exist along the $y$-direction.
Then the equations of motion at linear order simplify to\
$h_{ty}'' - {3\over r} h_{ty}' - k^2 h_{ty} = 0 ,\ \
h_{+y}'' - {3\over r} h_{+y}' - k^2 h_{+y} = - 2rh_{ty}'$,
from the $R_{-y}, R_{+y}$-equations respectively, which are now 
nontrivial. These equations at linear order are similar to 
(\ref{h++eom}) for the $g_{++}$-backgrounds we have discussed.
The solution\ $h_{ty}\sim r^2 I_2(kr)$\ that is normalizable at the 
boundary $r=0$ (with asymptotics\ $h_{ty}\sim r^4 ,\ r\ra 0$) grows 
in the interior as\ $h_{ty}\sim e^{kr}\ \ (r\ra\infty)$.
This $h_{ty}$-solution also implies a solution for $h_{+y}$.
These solutions exist for any $k>0$, and suggest striped phases in 
the IR (it is unclear at this level if there is any critical $k$). 
It can be checked that these solutions obtained at the 
linearised level are not solutions in the full system, which appears 
more complicated: the fact that these modes grow in the interior 
might suggest the breakdown of linear perturbation theory here, 
unlike the $g_{++}$-mode.
Likewise, the modes  \{$g_{+-}(x^+,x,r), \ g_{++}(x^+,x,r),\ g_{yy}(x,r)$\}
give rise to a closed subsystem of equations at the linearised level: 
the ansatz\
$g_{+-} = -{1\over r^2} (1 + h_{+-}(r) \sin kx) ,\
g_{++} = F (1 + h_{++}(r) \sin kx) ,\
g_{yy} = {1\over r^2} (1 + h_{yy}(r) \sin kx)$ ,
give at linear order the equations
$h_{yy}=-2h_{+-} ,\ h_{+-}'' - {3\over r} h_{+-}' - k^2 h_{+-} = 0 ,\
r^2 h_{++}'' + r h_{++}' - (4+k^2r^2) h_{++} = 4r h_{+-}'$.
As above, $g_{+-}$ satisfies the same equation as for a 
massless scalar.
These sorts of modes suggest different kinds of striped IR phases, 
albeit at linear order only, in asymptotically Lifshitz spacetimes, 
which are not null deformations. It would be interesting to explore 
these further.

\section{Holographic entanglement entropy}

We want to understand holographic entanglement entropy of subsystem
$A$ in the boundary field theory dual to an $AdS$ null deformation 
of the form (\ref{AdSnullLif}), using the prescription of \cite{Ryu:2006bv}
of finding the area of a bulk minimal surface bounding the subsystem $A$.
We will primarily discuss the case where subsytem $A$ has the shape 
of a strip in the $x,y$-plane. Since these systems are naturally 
described in terms of slicings at constant lightcone time $x^-$ 
(rather than a timelike coordinate), it is natural to consider a 
minimal surface at a constant-$x^-$ slice. The spatial metric on 
such a slice then is
\be
ds^2 = {R^2\over r^2} (dx_i^2 + dr^2) + g_{++}(r,x_i) (dx^+)^2\ ,
\ee
where we have reinstated the $AdS$ radius $R$.
Subsystems $A$ that straddle some part of the $x_i$-plane but extend 
along the $x^+$-direction completely are natural from the point of 
view of the lower dimensional theory obtained by dimensional reduction 
along the $x^+$-direction, as is implicit in \cite{Narayan:2012hk}: 
this then gives in the bulk a minimal surface wrapping the 
$x^+$-direction and bounding the subsystem $A$ in the $x_i$-plane.
If $g_{++}=0$, this surface degenerates and becomes null, as is 
clear from the spatial metric above.
Consider a strip region in the $x$-direction given by 
$-l\leq x\leq l$, extending along the $y$-direction: then we 
expect that the minimal surface is parametrized by $r=r(x)$, and its
area gives the entanglement entropy
\be\label{HolEE}
S_E = {1\over G_5} \int_0^L {Rdy\over r} \int_0^{L_+} \sqrt{g_{++}} dx^+ 
\int {R\sqrt{dx^2 + dr^2}\over r} \
=\ {L L_+R^2\over G_5} \int_\epsilon dr\ {\sqrt{g_{++}}\over r^2} 
\sqrt{1 + \Big({dx\over dr}\Big)^2}\ ,
\ee
where $\epsilon$ is the near-boundary cutoff (\ie\ the UV cutoff in 
the field theory). The minimal surface has a turning point $r_t$ 
where ${dr\over dx}|_{r_t}=0$. Let us now consider homogenous 
backgrounds, to begin with. Then $g_{++}$ has no $x_i$-dependence, 
giving the usual conserved conjugate momentum to $x(r)$, which 
enables us to solve for the minimal surface, its area and the 
entanglement entropy. For instance, for $z=2$ Lifshitz 
backgrounds, $g_{++}$ has no $r,x_i$-dependence: taking it to be 
constant for simplicity, we have\ 
$S_E\sim\ {L L_+\over G_5} \int_\epsilon dr\ {1\over r^2}\sqrt{1+(x')^2}$ .
After $x^+$-compactification, with ${G_5\over L}=G_4$, this is of 
the same form as the lower dimensional expression for a 4-dim 
Lifshitz spacetime (as well as $AdS_4$ with constant time slices).
The conserved momentum ${1\over r^2} {x'\over\sqrt{1+(x')^2}} = p$ 
gives the turning point $r_t={1\over\sqrt{p}}$,\ the equation 
for the surface\ ${dx\over dr} = {1\over\sqrt{1-p^2r^4}}-1$, thereby 
$l=r_t$ , and\ 
\be\label{EElif}
S_E\sim  {L L_+R^2\over G_5} 
\int_\epsilon^{r_t} dr\ {1\over r^2\sqrt{1-(r^4/r_t^4)}} \ \sim\ 
{L L_+R^2\over G_5 l} \int_{\epsilon/l}^1 du\ {1\over u^2\sqrt{1-u^4}}
\sim\ {L_+R^2\over G_5} \Big({L\over \epsilon} - \# {L\over l}\Big)\ ,
\ee
where the coefficient in the second term arises from\ 
$\int_0^1 du {1\over u^2} ({1\over\sqrt{1-u^4}}-1)$ which has a 
finite value. The first term is a reflection of the area law. This 
calculation is in fact technically the same as that for $AdS_4$ with 
constant time slices (see \eg\ \cite{Ryu:2006bv}).
It is worth emphasizing that this nonvanishing entanglement entropy
for the Lifshitz state is due to a nonvanishing $g_{++}$: using
constant-$x^-$ slices for the $AdS_5$ vacuum gives a vanishing result,
since $g_{++}=0$\ (while using slices of a timelike time coordinate
gives the usual expressions respecting the area law).

For the homogenous $AdS$ plane wave considered in \cite{Narayan:2012hk}, 
we have\ $g_{++}=R^2Qr^2$ , giving
\be
S_E =\ {L L_+R^3\sqrt{Q}\over G_5} \int_\epsilon^{r_t} dr\ 
{\sqrt{1 + (x')^2}\over r} \ ;\qquad\quad 
{1\over r} {x'\over\sqrt{1+(x')^2}} = p\ ,
\qquad l=r_t={1\over p}\ .
\ee
The minimal surface then is\ $x=l\sqrt{1-({r\over r_t})^2}$, and we have\
$S_E= {L L_+R^3\sqrt{Q}\over G_5}\ \int_{\epsilon/l}^1 {du\over u\sqrt{1-u^2}}$ , 
giving
\be\label{logviol}
S_E =\ {2 L_+R^3\sqrt{Q}\over G_5}\ L \log {l\over\epsilon}\ .
\ee
Using $G_4={G_5\over L_+}$ , this gives the logarithmic violation of 
the area law, as expected from the lower dimensional theory. We have 
effectively taken $l\gg \epsilon$, so that the strip width $l$ is 
macroscopic relative to the UV cutoff $\epsilon$ in the field theory. 
When the strip size shrinks to roughly the cutoff, we have a 
cross-over to the UV behaviour in the field theory: in this case, we 
expect the entanglement entropy for $AdS_5$ in lightcone time slicing 
which vanishes, as vindicated by (\ref{logviol}) for $l\sim\epsilon$.

Note that (\ref{EElif}) and (\ref{logviol}) both agree with the
corresponding expressions obtained from the lower dimensional
descriptions, in the the 4-dim $z=2$ Lifshitz spacetime and the
hyperscaling violating metric with $\theta=1, d=2$ respectively. The
latter was of course expected from the dimensional reduction in
\cite{Narayan:2012hk}. It is interesting that this calculation
(\ref{logviol}) above arises from just the 5-dim part of the
spacetime, so in particular it also applies to $AdS_5\times X^5$
homogenous plane waves dual to various \No\ super Yang-Mills theories.

Consider now an asymptotically Lifshitz solution superposed with the 
homogenous plane wave state\ $g_{++}=F+R^2Qr^2$. We then expect the 
entanglement entropy to exhibit a cross-over from Lifshitz ground 
state behaviour to the logarithmic violation above. 

Now let us consider inhomogenous backgrounds: for simplicity, we will 
focus here on backgrounds which preserve $y$-translation symmetry, 
\ie\ where $g_{++}$ has no $y$-dependence. Then consider a strip in 
the $x$-direction given by $-l\leq x\leq l$, extending along the 
$y$-direction: we expect that the minimal surface in this case is 
still parametrized by $r=r(x)$ on symmetry grounds. The entanglement 
entropy is then given by (\ref{HolEE}) as before.
If the strip region is symmetric w.r.t. the bulk geometry, we expect 
the minimal surface to have a turning point $r_t$ where 
${dr\over dx}|_{r_t}=0$. In general, regarding (\ref{HolEE}) as 
an action for $x(r)$, the Euler-Lagrange equation of motion for 
extremization gives\ ($x'\equiv {dx\over dr}$)
\be
{d\over dr} \left({\sqrt{g_{++}}\over r^2}\ {x'\over\sqrt{1+(x')^2}}\right)
= {\sqrt{1+(x')^2} \over 2r^2\sqrt{g_{++}}}\ {\del g_{++}\over\del x}\ .
\ee
Now consider a spacetime of the form (\ref{AdSnullLif}) where 
$g_{++}$ is a ``small'' inhomogenous departure from the homogenous 
$AdS$ plane wave, in the sense of (\ref{g++Smallxdep}), \ie\ with 
$T_{++}\sim Q+\epsilon f(x)$. Then we expect that at least for a 
strip of small width (and centered about the lump at $x=0$), the 
minimal surface will not dip into the interior too much and so will
not be sensitive to the horizon: this will again give the entanglement
entropy to be (\ref{logviol}) approximately. For larger strip widths,
the minimal surface will dip further into the interior and will 
exhibit significant departures. Perhaps analysing entanglement 
entropy further will give more insight into such backgrounds.


\section{Discussion}

We have described $AdS$ analogs of plane waves with possible
inhomogeneities. These are likely to be $\al'$-exact string
backgrounds. They correspond in the dual gauge theory to turning on
lightcone momentum density $T_{++}$ varying spatially. We have seen
that generically the inhomogenous $AdS$ plane waves exhibit analogs of
horizons where a timelike Killing vector becomes null. From the point
of view of $x^+$-dimensional reduction, the circle pinches off at
these horizon loci, string winding modes becoming light here,
indicating new stringy physics.

In general these are normalizable backgrounds (rather than
fluctuations) and grow in the interior, as we have discussed. One
might wonder if these $AdS$ plane waves with inhomogeneities are
unstable towards ``settling'' down to the homogenous one with uniform
$T_{++}$ studied in \cite{Narayan:2012hk}. However all these
backgrounds have finite curvature invariants, and preserve as much
supersymmetry as the homogenous case. It would be interesting to
understand this better, including finite temperature versions. In 
this regard, it is useful to recall that boosted black branes 
in lightcone variables \cite{Maldacena:2008wh} in a certain 
zero-temperature double scaling limit reduce to the homogenous 
$AdS$ plane wave, as discussed in \cite{Singh:2012un}. To elaborate, 
we see that the familiar black D3-branes\
$ds^2={1\over r^2} [-(1-r_0^4r^4) dt^2 + dx_3^2 + dx_i^2] 
+ {dr^2\over r^2 (1-r_0^4r^4)}$\ with the horizon at $r_0$ can be 
rewritten in lightcone variables\ $x^\pm={t\pm x_3\over\sqrt{2}}$\ 
after a boost by a parameter $\lambda$ as
\be
ds^2 = {1\over r^2} \left[-2dx^+dx^-+ {r_0^4 r^4\over 2} 
(\lambda dx^++\lambda^{-1} dx^-)^2 + dx_i^2\right] 
+ {dr^2\over r^2 (1-r_0^4r^4)}\ .
\ee
Now in the double scaling limit $r_0\ra 0,\ \lambda\ra\infty$, with 
$Q={r_0^4\lambda^2\over 2}$ fixed, we see that this reduces to 
(\ref{AdSnullLif}) with $g_{++}=Qr^2$, \ie\ the homogenous $AdS$ 
plane wave \cite{Narayan:2012hk}. We also note that the various 
energy-momentum components \cite{Maldacena:2008wh}
\be
T_{++}\sim \lambda^2 r_0^4\ ,\qquad 
T_{--}\sim {r_0^4\over\lambda^2}\ ,
\qquad T_{+-}\sim r_0^4\ ,\qquad T_{ij}\sim r_0^4\delta_{ij}\ ,
\ee
reduce to nonzero $T_{++}$ alone in this (highly chiral) limit. For
small $r_0$, we have a large left-moving wave with a small
right-moving sector. Now consider a small inhomogeneity in the finite
temperature gauge theory, dual to the black brane: it would appear
that this would have a corresponding bulk gravity description
(possibly more intricate) with inhomogeneities in the horizon. We
expect that this will equilibriate over time to give rise at late
times to a homogenous horizon. However under a boost $\lambda$, one
might think the timescale for this equilibriation might be dilated: 
in the extreme chiral limit $\lambda\ra\infty$, it might seem this 
timescale for ``settling'' down diverges, so that any initial
inhomogeneities prevail over a long timescale, thus suggesting the
inhomogenous backgrounds discussed here are quasi-stable in this sense.
This appears to dovetail with the supersymmetry preserved by all these
backgrounds. These arguments are quite heuristic of course: it is
worth understanding the stability of these backgrounds better.

We have also described asymptotically Lifshitz backgrounds of this
sort with possible inhomogeneities, with some discussion of Lifshitz
singularities from the perspective of the $AdS/CFT$ construction
involving null deformations. The Lifshitz vacuum appears to be a
non-generic state in these constructions. It would be interesting to
obtain a deeper understanding of string propagation in such
backgrounds, as well as the rules of $AdS/CFT$ and its
deformations/states for such backgrounds.

Finally it is interesting to explore holographic entanglement entropy
in these backgrounds, especially from the point of view of the dual
field theory. We recall that the homogenous background (for $AdS_5$)
with uniform lightcone momentum density \cite{Narayan:2012hk}
interestingly exhibits logarithmic violation of the area law of
entanglement entropy holographically. We hope to explore this 
further \cite{wip}.

\vspace{10mm}
\noindent {\small {\bf Acknowledgments:} It is a pleasure to thank 
K. Balasubramanian, S. Das, G. Horowitz, N. Iqbal, S. Kachru, 
S. Minwalla, M. Rangamani, E. Silverstein, N. Suryanarayana, T.
Takayanagi and S. Trivedi for helpful conversations on various aspects
of this work. I also thank the hospitality of the KITP Santa Barbara,
Stanford Institute for Theoretical Physics, Physics Dept, U. Kentucky,
USA, the Organizers of the String Theory Discussion Meeting, June '12, 
at the International Center for Theoretical Sciences (ICTS), Bangalore, 
and the String Theory groups at TIFR, Mumbai, at various stages while 
this work was being carried out. This work is partially supported by 
a Ramanujan Fellowship, DST, Govt. of India.}

\end{document}